\documentclass{PoS}

\title{ProtoDUNE and a Dual-Phase LArTPC}

\ShortTitle{ProtoDUNE and a Dual-Phase LArTPC}

\author{\speaker{Andrea Scarpelli}\thanks{ This project has received funding from the European Union's Horizon 2020 research and innovation programme under the Marie Sklodowska-Curie grant agreement No 665850. } ,on behalf of the DUNE Collaboration
\\
        APC, Universit\'e Paris Diderot, CNRS/IN2P3, CEA/Irfu, Obs de Paris, USPC, Paris 75205, France.\\
        E-mail: \email{scarpell@apc.in2p3.fr}}

\abstract{The four 10 kt Liquid Argon Time Projection Chambers (LArTPCs) of the future DUNE experiment will enable precise measurements of the oscillation parameters and the discovery of CP violation for leptons, thanks to their excellent 3D imaging capabilities and calorimetric capabilities. One or more modules of the DUNE detector may exploit a Dual Phase (DP) LArTPC that, relying on the extraction of the charge produced in the liquid volume and its subsequent multiplication in argon gas, may offer a robust and competitive signal-to-noise ratio and a fully active volume. In 2018 and 2019, the ProtoDUNE experiment at CERN will validate the  designs of the DUNE far detector, showing the feasibility of large scale Dual-Phase LArTPC and providing precious insight on the DUNE physics potential.}

\FullConference{Neutrino Oscillation Workshop (NOW2018)\\
		9 - 16 September, 2018\\
		Rosa Marina (Ostuni, Brindisi, Italy)}

\begin{document}

\section{Introduction}
The next generation (>2020) of long-baseline neutrino experiments, such as DUNE \cite{dunecdr}, will perform measurements of the oscillation parameters, determine the Mass Ordering and search for CP violation in the leptonic sector with unprecedented precision. In order to achieve these ambitious goals, the DUNE experiment will deploy four massive Liquid Argon Time Projection Chambers (LArTPCs), for a total fiducial mass of 40 kt \cite{dunecdr}. LArTPCs allow to build large high-density homogeneous calorimeters with a high resolution ($\sim\:$mm) tracking performance. Since Argon is also transparent to its own light, prompt scintillation photons may be used as the event trigger. Understanding the detector response and proving the feasibility of this technology at the kt scale is of great importance for the DUNE physics program.

\section{LArTPC at the scale of DUNE}
 In LArTPCs, 3D images of the event can be formed by collecting the electrons produced during the Argon ionization and transported onto a readout thanks to a high drifting field. During the drift, some electrons can be trapped by impurities in the Argon volume. The number of survived electrons after a given drift time follows an exponential decay whose lifetime $\tau$ depends on the Argon purity. Assuming $\tau\,=\,3.0$ ms \footnote{ corresponding to an Argon purity of $\approx$300$\;$ppt/ms $O_2$ \cite{pdunedptdr}} and a drift velocity of $1.6$\,mm/$\mu$s, the amount of charge loss after a 3 meters-long drift is 20$\,\%$ and of about 93$\,\%$ for drift distances at the scale required for DUNE ( $\sim$ 12$\,$m ). 

\begin{figure}
\centering
\includegraphics[width=0.6\textwidth]{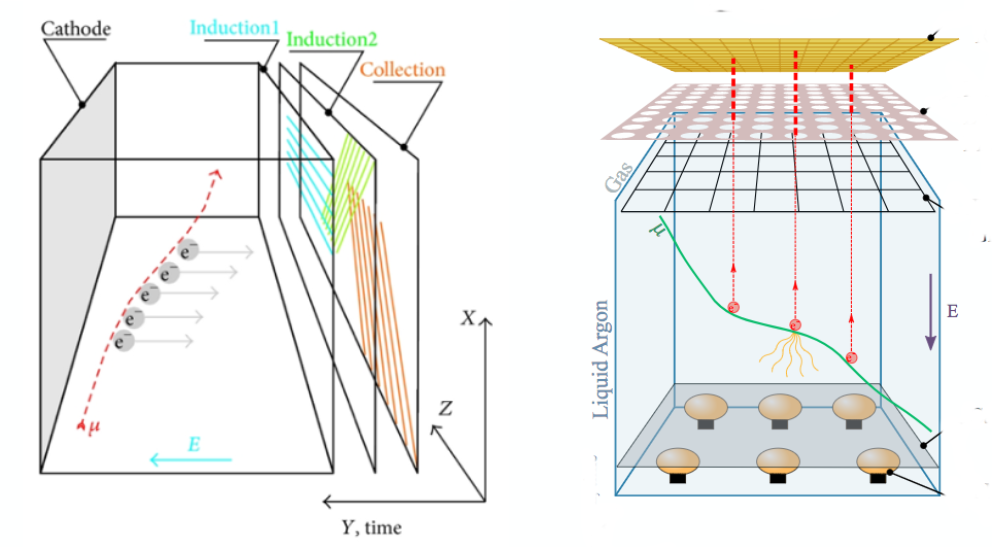}
\caption{Sketch of two possible setups considered for the DUNE far detector LArTPC. Left: Single-Phase setup. Right: Dual-Phase setup.}
\label{fig:setup}
\end{figure}

 In a Single-Phase (SP) LArTPC (Figure \ref{fig:setup} left ) electrons are drifted horizontally onto a readout composed of two induction and one collection wire planes. Although this setup is the most common for this type of detectors, drift length cannot exceed 3-4 meters and therefore using this techonolgy for DUNE requires a complex modular design with several TPCs within the Ar volume.

\begin{figure}
\centering
\includegraphics[width=0.8\textwidth]{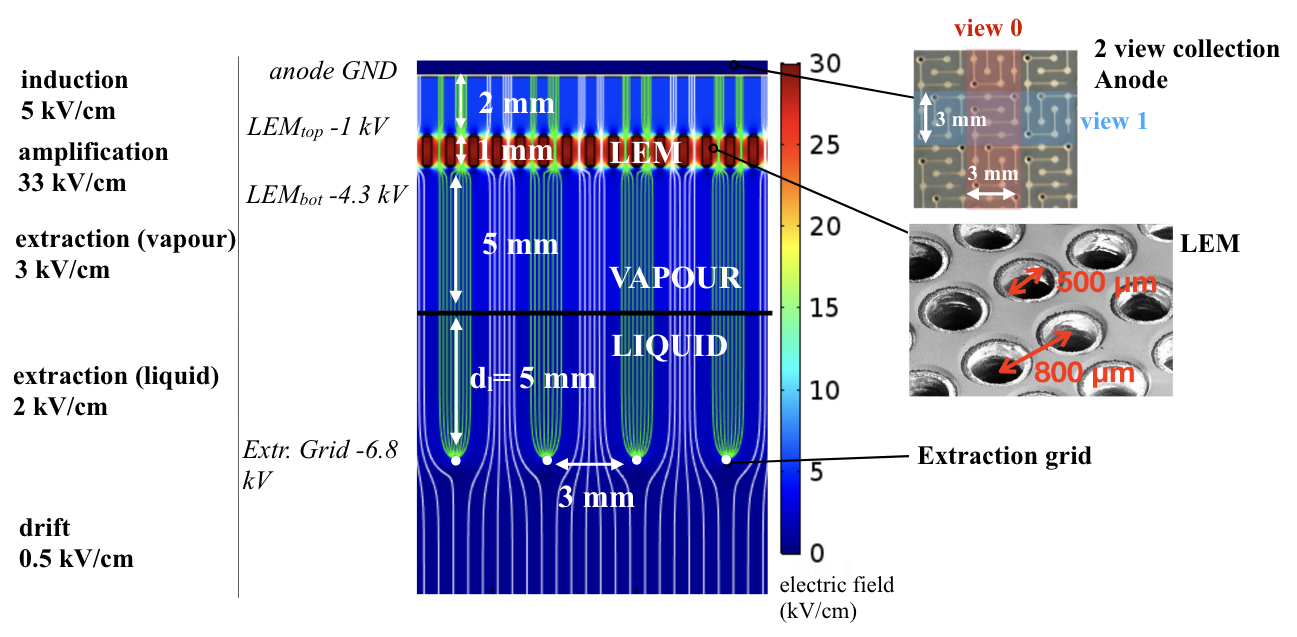}
\caption{Left: illustration of the extraction, amplification and readout regions in a dual-phase LArTPC. The simulated  field  lines  in  white  are  an  indication  of  those  followed  by  the drifting  charge. The quoted electric fields (in bold) and electrode potentials (in italic) correspond to a stable effective gain of around 20 \cite{311}}
\label{fig:sandwich}
\end{figure}

One alternative approach is a Dual-Phase (DP) setup, in Figure \ref{fig:setup} right. In this case the electrons are drifted vertically and amplified by a strong filed applied in a thin gas Argon layer just before the readout. The amplification can be tuned to compensate for electron losses or high noise,  enabling fully active drift distances of several meters. The Dual-Phase design of the DUNE far detector features a modular readout composed of 20 Charge Readout Planes (CRP) located at the surface of the Argon volume with 36 LEM-Anode assemblies of 50x50 cm$^2$ each. The CRPs work \textit{sandwich} as shown in figure \ref{fig:sandwich}: electrons are extracted from liquid into gas with an efficiency higher than 90$\:\%$ and amplification is operated by a 1 mm thick Large Electron Multipliers (LEMs). The anode is composed of two collection planes designed to maximize the readout granularity (3 mm), while ensuring an equal charge sharing and noise within 1500 ENC. Since the readout is located at the top of the Argon volume, cold FE electronics remains accessible and may be easily replaced in case of failure.

A 4 t fiducial mass demonstrator of this technology, with a single 1x3 m$^2$ CRP,  1 m drift has been operated at CERN in 2017 with cosmic rays, demonstrating the overall detector concept \cite{311}.

\section{The ProtoDUNE experiment}
The ProtoDUNE experiment at CERN has the main goal to validate both the SP and DP design for the DUNE far detector \cite{dunecdr} and provide insight on the detector response and calibration with both a charge particle beam and cosmic rays. It also aims to perform measurements of the Ar-$\pi$ cross section in energy ranges critical for DUNE. Both ProtoDUNEs are supposed to test the proposed design of all the component at 1:1 scale wherever it is possible, aiming at demonstrating the long-term stability of the detector operations.

\begin{figure}
\centering
\includegraphics[width=0.8\textwidth]{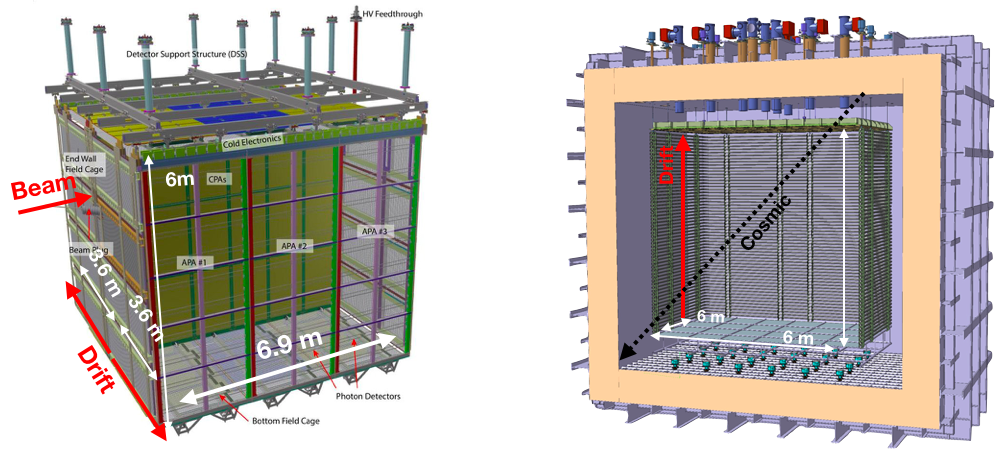}
\caption{Left: draft of ProtoDUNE-SP \cite{pdunesptdr}. Right: draft of ProtoDUNE-DP \cite{pdunedptdr}}
\label{fig:protodune}
\end{figure}

ProtoDUNE-SP (Figure \ref{fig:protodune} left) features total Ar mass of 0.7 kt and a total drift distance of 7.2 m, split into two 3.6 m sections. The central cathode provides the drift field necessary to push the electrons on the Anode Plane Assemply (APA) located at the edge of the detector. Each APA is composed by 2 induction and 1 collection plane with a 5 mm wire pitch. The FE electronics and the photodetection system are both integrated in the APA.

ProtoDUNE-DP (Figure \ref{fig:protodune} right) has an active mass of 0.3 kt and a total drift distance of 6 m (one-half of the total drift at DUNE). The readout includes 4 DUNE-CRPs with 36 50x50 m$^2$ LEMs each. The photodetection system is composed of 36 PMTs located at the bottom of the detector.

\section{Conclusion}
Intense R$\&$D is ongoing in order to prove the reliability and scalability of LArTPC technologies for the next generation of long-baseline neutrino experiment such as DUNE. The ProtoDUNE experiment is conceived to validate both the Single and Dual Phase design proposed for DUNE and provide important insight on the DUNE physics potential. The assembly of ProtoDUNE-SP was completed during Summer 2018 and the detector took both beam and cosmic data during Fall 2018. The assembly of ProtoDUNE-DP is still ongoing at CERN and plans to collect cosmic ray data in Summer 2019.

\end{document}